\documentclass[cameraready]{Interspeech}

\usepackage{mdframed}
\usepackage[most]{tcolorbox}
\usepackage{multirow}
\usepackage{subcaption}

\title{Ecologically-Constrained Task Arithmetic for Multi-Taxa Bioacoustic Classifiers Without Shared Data}

\author[affiliation={1}, orcid=0000-0002-3876-3033]{Ragib Amin}{Nihal}
\author[affiliation={1,2}, orcid=0000-0002-6958-7319]{Benjamin}{Yen}
\author[affiliation={1}]{Runwu}{Shi}
\author[affiliation={1}]{Takeshi}{Ashizawa}
\author[affiliation={1}]{Kazuhiro}{Nakadai}

\address{
    $^1$ Systems and Control Engineering, Institute of Science Tokyo, Japan \\
    $^2$ RIKEN BDR, Japan
}

\email{ragib@ra.sc.e.titech.ac.jp}


\usepackage{comment}

\begin{document}

\maketitle

\begin{figure*}[!t]
\centering
\includegraphics[width=0.85\textwidth]{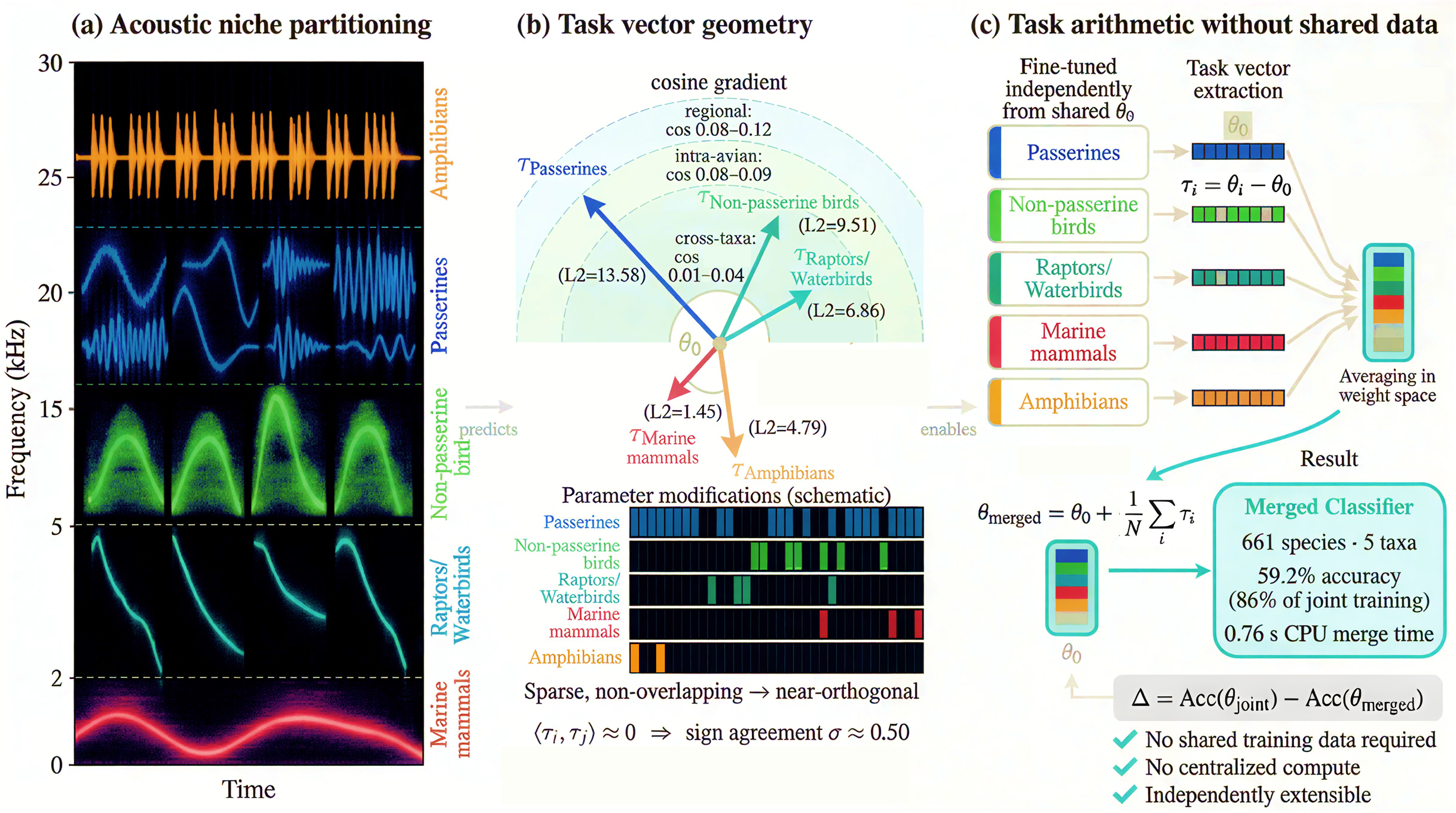}
\vspace*{-0.3cm}
\caption{\footnotesize{Ecologically-constrained task arithmetic for bioacoustic model composition. \textbf{(a)}~Acoustic niche partitioning: taxonomic groups concentrate vocal energy in non-overlapping frequency bands (schematic). \textbf{(b)}~Top: task vectors are near-orthogonal in weight space, with magnitude proportional to dataset size. Bottom: each group modifies a sparse, largely disjoint subset of encoder parameters. \textbf{(c)}~Composition pipeline: independently fine-tuned specialists yield task vectors that are averaged and added to the shared pretrained encoder, producing a unified classifier without shared data.}}

\label{fig:overview}
\vspace*{-.5cm}
\end{figure*}
\begin{abstract}
Training data for bioacoustics is scattered across taxa, regions, and institutions. Centralizing it all is often infeasible. 
We show that independently fine-tuned BEATs encoders can be composed into a unified 661-species classifier via task vector arithmetic without sharing data. 
We find that bioacoustic task vectors are near-orthogonal (cosine 0.01--0.09). Their separation aligns closely with spectral distribution distance, a gradient consistent with the acoustic niche hypothesis.
This geometry makes simple averaging optimal while sign-conflict methods reduce accuracy by one to six percentage points. Composition also creates an asymmetric gap: species-rich groups lose accuracy relative to joint training while underrepresented taxa gain, a redistribution useful for equitable biodiversity monitoring. We verify linear mode connectivity across all taxonomic pairs, demonstrate zero-shot transfer to new regions, and identify domain negation as a boundary condition where composition fails. These results enable a collaborative paradigm for bioacoustics where institutions share only task vectors to assemble multi-taxa classifiers, preserving data privacy.
\href{https://github.com/Ragib-Amin-Nihal/BioAcousticArithmetic}{\textcolor{blue}{\underline{Code Link}}}
\end{abstract}

\section{Introduction}
\label{sec:intro}

Passive bioacoustic monitoring generates large volumes of recordings across thousands of sites, capturing sounds from birds, marine mammals, amphibians, and insects~\cite{stowell2022computational}. However, the training data needed to build automated species classifiers remain fragmented. Ornithological surveys, cetacean programs, and herpetological fieldwork each use different equipment, annotation standards, and data-sharing policies~\cite{kahl2021birdnet,sayigh2016watkins,canas2023anuraset}. Geographic coverage is similarly uneven: the BirdCLEF competition series rotates annually across East Africa, South Asia, and the Neotropics, producing datasets for species sets that have minimal overlap.
\\
Monolithic classifiers such as BirdNET~\cite{kahl2021birdnet} and Perch~\cite{williams2024perch} have made monitoring more accessible, but extending them to new taxa or regions requires full retraining. Fine-tuning offers a more efficient alternative by adapting pretrained encoders to local contexts~\cite{denton2025agile}, yet each resulting model remains isolated. A classifier trained on cetaceans and one trained on passerines cannot be combined without joint retraining or suffering from catastrophic forgetting~\cite{french1999catastrophic}. Training all tasks jointly achieves high accuracy but demands centralized data access and complete retraining whenever new data are added. Currently, no method allows independently trained specialists to pool their knowledge without access to the original training data.
\\
Task arithmetic provides a promising solution~\cite{ilharco2023editing}. Starting from a pretrained encoder $\boldsymbol{\theta}_0$ and a model fine-tuned on task $A$ with parameters $\boldsymbol{\theta}_A$, the difference $\boldsymbol{\tau}_A = \boldsymbol{\theta}_A - \boldsymbol{\theta}_0$ forms a task vector representing what was learned. Adding task vectors from different specialists reconstructs a multi-task model without combined training. This approach has been validated in computer vision~\cite{ilharco2023editing, yadav2023ties} and is beginning to be explored in speech processing~\cite{rittergutierrez2025audio}, but it has not been tested in bioacoustics. This paper investigates the \textbf{research question} of:

\begin{mdframed}[
  leftline=true,
  rightline=false,
  topline=false,
  bottomline=false,
  linewidth=2pt,
]
\textit{whether independently fine-tuned bioacoustic classifiers can be composed via task vector arithmetic and whether ecological principles can predict the geometry of the resulting weight-space structure?}
\end{mdframed}

\noindent We hypothesize that the acoustic niche hypothesis~\cite{krause1993niche} may extend to transformer weight space. If species partition spectro-temporal space to reduce masking (Fig.~\ref{fig:overview}a), then classifying taxonomically distinct groups would modify largely disjoint parameter subsets, leading to near-orthogonal task vectors (Fig.~\ref{fig:overview}b) that can be composed without interference (Fig.~\ref{fig:overview}c). Two consequences follow, derived formally in \S\ref{sec:theory}. (i)~sign-conflict resolution methods like TIES~\cite{yadav2023ties} become harmful rather than helpful. Near-chance sign agreement makes majority-vote election random; (ii)~uniform averaging produces asymmetric composition gaps that favor minority taxa over majority taxa. In audio, Ritter-Gutierrez et al.~\cite{rittergutierrez2025audio} found that TIES fails for speech-music merging but offered no geometric explanation. Marincione et al.~\cite{marincione2025naturelm} applied simple interpolation to a bioacoustic audio-language model without examining multi-task composition or task vector geometry. No prior work has characterized bioacoustic task vector geometry or verified linear mode connectivity for any audio encoder. We address both across taxonomic composition (five groups, 661 species), geographic composition (four regions), and focal-to-soundscape domain negation; also correlating pairwise spectral distribution distances with task vector cosine similarities.
Contributions: \\
\textbf{(1) A composition framework for multi-taxa bioacoustic classifiers. }~We show that independently fine-tuned species-group models compose into a 661-species classifier achieving 59.2\% accuracy (86\% of a jointly-trained baseline) without shared data or compute, with regional models achieving 91\% of dedicated performance. \\
\textbf{(2) Evidence for weight-space geometry.}~We find that bioacoustic task vectors are near-orthogonal (cosine similarities 0.01--0.09), with similarities increasing from cross-taxa pairs (0.01--0.04) to intra-avian pairs (0.08--0.09) to regional pairs (0.08--0.12). This gradient is strongly correlated with spectral distribution distance, a relationship consistent with the acoustic niche hypothesis.
\\\textbf{(3) Discovery of asymmetric composition effects that benefit underrepresented taxa.}~Merging redistributes capacity from majority to minority groups: passerines lose 11.8\% while marine mammals \textit{gain} 3.9\% and amphibians gain 1.9\%.
\vspace*{-0.2cm}
\section{Method}
\label{sec:method}
\vspace*{-0.2cm}
\subsection{Problem setting}
\label{sec:setting}
\vspace*{-0.2cm}
We consider building a multi-taxa classifier from $N$ independently trained specialists without access to their training data. Given a shared pretrained encoder $\boldsymbol{\theta}_0 \in \mathbb{R}^d$ and $N$ specialist checkpoints $\{\boldsymbol{\theta}_1, \ldots, \boldsymbol{\theta}_N\}$, each fine-tuned on a private dataset $D_i$ covering a disjoint species set $S_i$ ($S_i \cap S_j = \emptyset$ for $i \neq j$), we define the \textit{task vector} $\boldsymbol{\tau}_i = \boldsymbol{\theta}_i - \boldsymbol{\theta}_0$ as the direction learned by specialist $i$. The objective is to produce a single encoder $\boldsymbol{\theta}_{\mathrm{merged}}$ that classifies species from $\bigcup_i S_i$ using only the checkpoints and base model, no training data is exchanged.
\\
This reflects common constraints (detailed in Appendix~\ref{app:scenarios}): data cannot be shared due to agreements or sensitivity; archives are too large to centralize (terabytes per year)~\cite{nihal2025weakly}; new taxa emerge after deployment; regions lack local data.
\\
\textbf{Assumptions.} Species sets are disjoint; all specialists fine-tuned from the same $\boldsymbol{\theta}_0$ with identical hyperparameters; each specialist contributes only its task vector $\boldsymbol{\tau}_i$. The second is binding but requires only lightweight coordination (e.g., agreeing on a base model and config). Since $\boldsymbol{\tau}_i$ always has $d$ parameters, composition cost is independent of training data size.

\vspace*{-0.4cm}
\subsection{System overview}
\label{sec:pipeline}
\vspace*{-0.2cm}

\textit{Stage~1: Independent fine-tuning.} After preprocessing, each group fine-tunes $\boldsymbol{\theta}_0$ on $D_i$, producing $\boldsymbol{\theta}_i$.
\\
\textit{Stage~2: Task vector extraction and merging.} Computed on encoder weights only (excluding classification heads), merged via the strategies in \S\ref{sec:merging}, added to $\boldsymbol{\theta}_0$ to produce merged encoder.
\\
\textit{Stage~3: Evaluation.} The frozen merged encoder is evaluated via linear probing: a single linear layer is trained on the encoder's output features. We train both per-group probes (within-group accuracy) and a unified probe over all species (multi-taxa classification). $k$-Nearest Neighbor ($k{=}1$) provides a probe-independent diagnostic. The primary metric is the composition gap, comparing the merged model to a jointly-trained baseline. Accuracy differences are assessed with paired bootstrap 95\% Confidence Intervals (10{,}000 resamples).

\vspace*{-0.2cm}
\subsection{Formulation and analysis}
\label{sec:theory}
\vspace*{-0.2cm}
The simplest composition averages task vectors:
%
\begin{equation}
  \boldsymbol{\theta}_{\mathrm{merged}} = \boldsymbol{\theta}_0 + \tfrac{1}{N} \textstyle\sum_i \boldsymbol{\tau}_i.
  \label{eq:merge}
\end{equation}
The composition gap $\Delta = \mathrm{Acc}(\boldsymbol{\theta}_{\mathrm{joint}}) - \mathrm{Acc}(\boldsymbol{\theta}_{\mathrm{merged}})$ measures the cost of composition versus joint training, both evaluated via linear probing on frozen encoders.


\noindent The merged encoder displaces specialist by $\boldsymbol{\delta}_i = \boldsymbol{\theta}_{\mathrm{merged}} - \boldsymbol{\theta}_i$. When task vectors are pairwise orthogonal, the displacement:
%
\begin{equation}
||\boldsymbol{\delta}_i||^2 = \tfrac{(N{-}1)^2}{N^2} ||\boldsymbol{\tau}_i||^2 + \tfrac{1}{N^2} \textstyle\sum_{j \neq i} ||\boldsymbol{\tau}_j||^2.
\label{eq:displacement}
\end{equation}
%
\noindent Under orthogonality, no directional interference occurs. Performance degradation depends only on vector magnitudes. Because the dominant term scales with $||\boldsymbol{\tau}_i||^2$, groups trained on larger datasets experience greater dilution, leading to asymmetric accuracy gaps. For sign conflict resolution methods, pairwise sign agreement 
$
\sigma(\boldsymbol{\tau}_i, \boldsymbol{\tau}_j)
=
\frac{1}{d}
\bigl|
\{\, k : \mathrm{sign}(\tau_i^k) = \mathrm{sign}(\tau_j^k) \,\}
\bigr|
$
approaches 0.5 under orthogonality. TIES-Merging's majority vote becomes random, discarding half of each vector's parameters. From this we can have the two prediictions discussed in~\S\ref{sec:intro}.
\vspace*{-0.3cm}
\subsection{Merging strategies}
\label{sec:merging}
\vspace*{-0.2cm}
Existing merging methods fall into two families: \textit{Direct combination} averages task vectors with optional rescaling. \textit{Conflict resolution}, which identifies parameters where task vectors disagree in sign and resolves disagreements before merging. 
\\
We evaluate three direct methods: \textbf{simple averaging} ($\frac{1}{N}\sum_i \boldsymbol{\tau}_i$); \textbf{task arithmetic}~\cite{ilharco2023editing} ($\lambda \sum_i \boldsymbol{\tau}_i$, $\lambda \in \{0.1, \ldots, 1.0\}$); and \textbf{DARE}~\cite{yu2024dare} (random dropout at rate $p$ with rescaling to preserve expectation; $p \in \{0.5, 0.7, 0.9, 0.95, 0.99\}$). 
\\
We also evaluate three conflict-resolution methods: \textbf{TIES}~\cite{yadav2023ties} (trim smallest values, elect sign by majority vote, merge; $k \in \{0.1, 0.2, 0.5, 0.8\}$); \textbf{DARE+TIES} (DARE dropout followed by TIES sign election); and \textbf{DELLA}~\cite{panigrahi2024della} (magnitude-proportional dropout replacing DARE's uniform sampling, followed by TIES). Algorithmic details in Appendix~\S\ref{app:merging_background}.
\vspace*{-0.2cm}
\subsection{Data and training}
\label{sec:setup}

\textbf{Model.} We use BEATs (iter3+ AS2M)~\cite{chen2023beats}, a 90-million parameter audio spectrogram transformer pretrained on AudioSet-2M through iterative self-supervised learning. BEATs uses LayerNorm, so merged encoders do not require recalibration.
\\
\textbf{Data.} We partition 661 species into five taxonomic groups (Table~\ref{tab:data}) drawn from BirdCLEF~23/24/25~\cite{kahl2021birdnet}, the Watkins Marine Mammal Sound Database~\cite{sayigh2016watkins}, and AnuraSet~\cite{canas2023anuraset}. Groups follow eBird taxonomy. 
Full taxonomic lists are in Appendix~\ref{app:datasets}. For regional experiments (\S\ref{sec:regional}), we use four geographic subsets: R1 BirdCLEF 2023 (East Africa), R2 BirdCLEF 2024 (South Asia), R3 BirdCLEF 2025 (Neotropics), and R4 BirdSet POW (North America). Each species is split 70/10/20 (train/val/test).

\begin{table}[t]
\centering
\caption{Species-group datasets and task vector properties. Sparsity: fraction of parameters with $|\tau| < 10^{-3}$.}
\vspace*{-.2cm}
\label{tab:data}
\small
\resizebox{\linewidth}{!}{
\setlength{\tabcolsep}{3.5pt}
\begin{tabular}{@{}llrrr|rrr@{}}
\toprule
 & Scope & Classes & Train & Source & $||\boldsymbol{\tau}||_2$ & Mean $|\tau|$ & Sparse \\
\midrule
G1 & Passerines       & 336 & 81k  & BirdCLEF & 13.58 & 1.05e-3 & 59.7\% \\
G2 & Non-pass.\ birds & 157 & 38k  & BirdCLEF &  9.51 & 7.46e-4 & 72.8\% \\
G3 & Raptors/waterb.  &  84 & 21k  & BirdCLEF &  6.86 & 5.43e-4 & 84.3\% \\
G4 & Marine mammals   &  21 & 1.4k & Watkins  &  1.45 & 1.22e-4 & 100\%  \\
G5 & Amphibians       &  63 & 12k  & AnuraSet &  4.79 & 3.84e-4 & 94.2\% \\
\bottomrule
\end{tabular}
}
\vspace*{-.5cm}
\end{table}

\noindent\textbf{Fine-tuning.} Each group independently fine-tunes the full BEATs encoder on its own dataset, starting from the same pretrained checkpoint. All runs use identical hyperparameters: AdamW (lr\,$=$\,$10^{-5}$, wd\,$=$\,0.01), OneCycleLR, batch~32, 20 epochs (patience~5), BF16, SpecAugment, Mixup ($\alpha{=}0.3$), label smoothing ($\varepsilon{=}0.1$). Task vectors are computed on encoder weights only, excluding the per-group classification heads.
\vspace*{-0.2cm}
\section{Experiments and Results}
\label{sec:experiments}
\vspace*{-0.1cm}
\subsection{Experiment 1: Linear Mode Connectivity}
\label{sec:lmc}
\begin{figure}[t]
\centering
\includegraphics[width=\columnwidth]{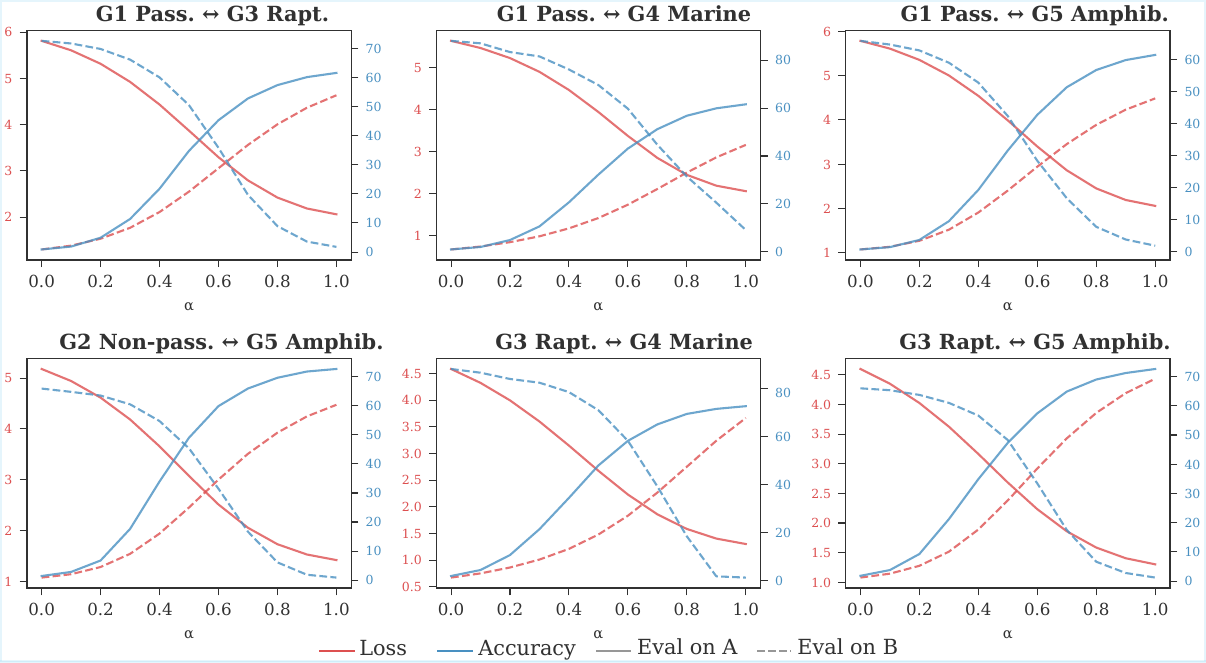}
\vspace*{-0.4cm}
\caption{Linear mode connectivity for some specialist pairs. Every curve is monotonic: no loss barrier exceeds endpoint.}
\label{fig:lmc_curves}
\vspace*{-0.4cm}
\end{figure}

\vspace*{-0.2cm}
Task vector addition relies on the assumption that all fine-tuned models lie in the same loss basin with $\boldsymbol{\theta}_0$. If this does not hold, merging could produce a model with high loss. We test this prerequisite for all ten pairwise combinations of group encoders by interpolating between each pair. For two specialists $\boldsymbol{\theta}_A$ and $\boldsymbol{\theta}_B$, we evaluate models along the path $\boldsymbol{\theta}(\alpha) = \alpha \boldsymbol{\theta}_A + (1-\alpha)\boldsymbol{\theta}_B$ at $\alpha \in \{0, 0.1, \ldots, 1.0\}$. We measure loss barriers using the corrected metric from Frankle et al.~\cite{frankle2020lmc}: $\max(0,\, \max_{\alpha \in (0,1)} L_\alpha - \max(L_0, L_1))$. Each pair is evaluated on the test sets of both endpoint groups.
\\
Every interpolation curve is monotonic (Fig.~\ref{fig:lmc_curves}). As $\alpha$ moves from one specialist toward other, loss changes smoothly without any intermediate point exceeding both endpoints. Midpoint accuracies at $\alpha=0.5$ range from 32\% to 71\%. This reflects the expected cost of interpolating between encoders specialized for different tasks, not any barrier violation. Universal LMC confirms that all fine-tuned models share a single loss basin. 

\vspace*{-0.3cm}
\subsection{Experiment 2: Species-Group Composition}
\label{sec:composition}
\begin{figure}[t]
\centering
\includegraphics[width=0.5\columnwidth]{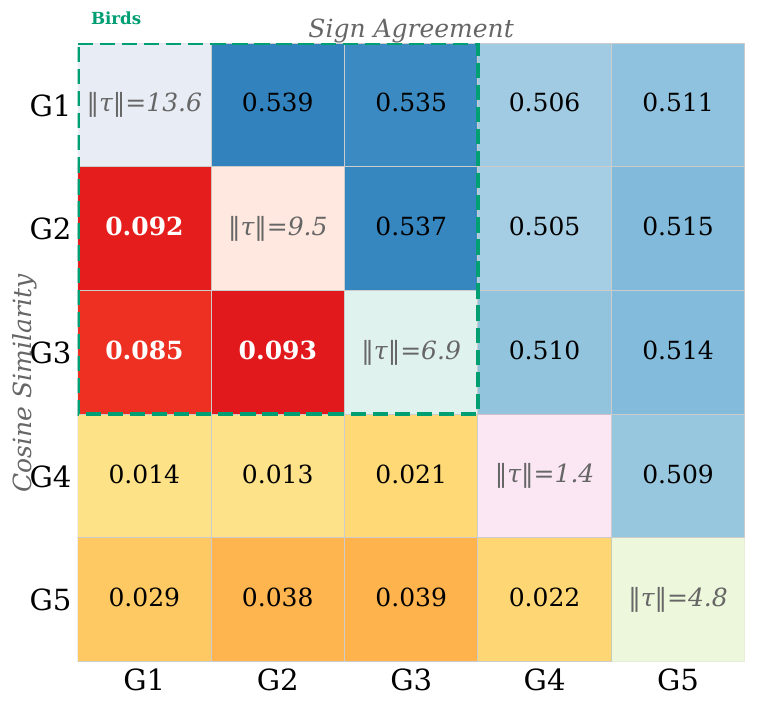}
\vspace*{-0.4cm}
\caption{Pairwise cosine similarity heatmap. 
}
\label{fig:cosine_matrix}
\vspace*{-0.5cm}
\end{figure}

\vspace*{-0.2cm}
Having verified that all fine-tuned models reside in the same loss basin, we next characterize the geometry of their task vectors to test the predictions of the acoustic niche hypothesis.
\\
\textbf{Task vector geometry:}
Two patterns appear in the pairwise cosine similarity heatmap (Fig.~\ref{fig:cosine_matrix}): Intra-avian pairs have cosine values between 0.085 and 0.093, with sign agreement from 0.535 to 0.539. Cross-taxa pairs are lower, ranging from 0.013 to 0.039 for cosine and 0.506 to 0.515 for sign agreement. These values are substantially below those reported for vision task vectors. In vision, semantically related tasks retain enough similarity for sign-conflict methods to help.
$\text{intra-avian} > \text{bird} \leftrightarrow \text{amphibian} > \text{bird} \leftrightarrow \text{marine mammal}$ 
pairs follow the same pattern as acoustic relatedness rather than factors unrelated to the tasks. Task vector L2 norms span from 13.58 for G1 to 1.45 for G4, correlating with training set size (Table~\ref{tab:data}). Per-layer analysis shows all groups increase modifications toward later transformer layers (Fig.~\ref{fig:layer_heatmap}, Appendix~\S\ref{app:per_layer}).

\noindent\textbf{Spectral distribution distance:} To test 
whether spectral similarity predicts task vector geometry, we correlated spectral distance with task vector cosine similarity over all pairs. Spectral distance was measured as Jensen-Shannon divergence on mean log-mel profiles using 500 clips per group. The correlation is strong and negative: Spearman $\rho = -0.915$ (permutation $p < 0.001$, $10^5$ permutations). Three clusters: bird-bird, bird-amphibian, and marine mammal-involved pairs correspond to graded acoustic distance. Mean-centering profiles to remove recording level confirms spectral shape drives the relationship ($\rho = -0.842$, $p = 0.004$). Excluding the four G4 (hydrophone) pairs, the trend persists ($\rho = -0.771$, $n{=}6$) but lacks power for conventional significance (Appendix~\S\ref{app:spectral}).
\begin{tcolorbox}[colback=gray!10, colframe=black, boxrule=1.0pt, arc=1mm,left=0.5mm, right=0.5mm, top=0.5mm, bottom=0.5mm]
\centering
{\small{\textbf{F1:} Bioacoustic task vectors are near-orthogonal, with geometry predicted by spectral distribution distance.}}
\end{tcolorbox}


\begin{figure}[t]
\centering

\begin{subfigure}[t]{0.42\columnwidth}
    \centering
    \includegraphics[width=\linewidth]{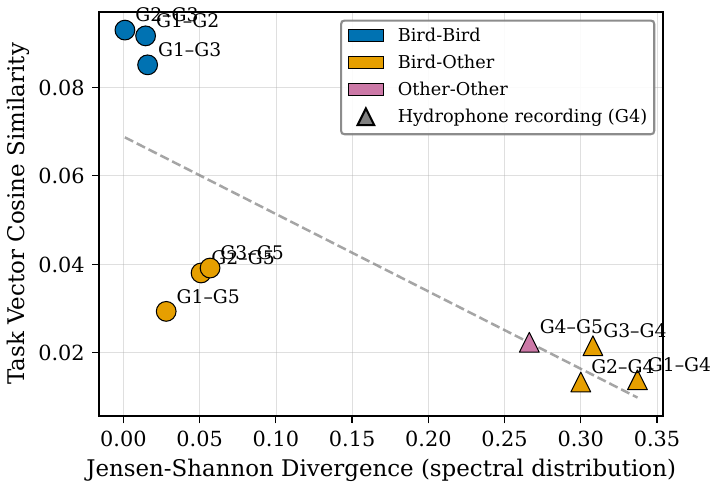}
    \caption{}
    \label{fig:spectral_vs_cosine}
\end{subfigure}
\hfill
\begin{subfigure}[t]{0.57\columnwidth}
    \centering
    \includegraphics[width=\linewidth]{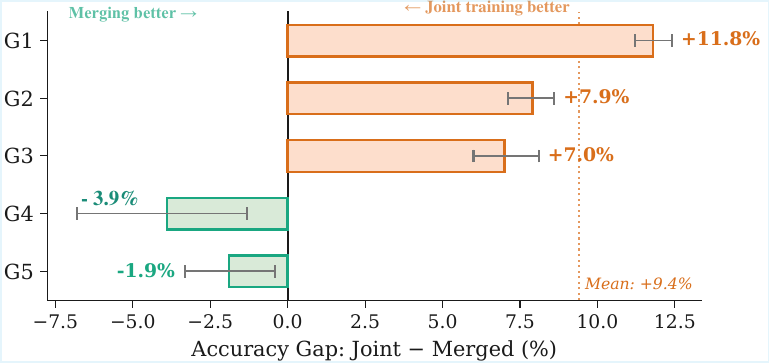}
    \caption{}
    \label{fig:asymmetric_gap}
\end{subfigure}
\vspace*{-0.4cm}
\caption{%
(a) Spectral distribution distance (JSD) vs.\ task vector cosine similarity. 
(b) Per-group composition gap. 
}
\label{fig:combined_analysis}

\vspace*{-0.2cm}
\end{figure}

\noindent\textbf{Method comparison:} Table~\ref{tab:methods} ranks the six merging methods. The top three methods fall within 0.35\% of each other. Every method that incorporates sign-conflict resolution underperforms simple averaging, confirming Prediction 1 from \S\ref{sec:theory}.


\begin{table}[t]
\centering
\caption{Merging methods ranked by 661-class accuracy. Joint-trained baseline: 68.3\%. $\dagger$ indicates sign-conflict resolution.}
\label{tab:methods}
\vspace*{-0.2cm}
\small
\setlength{\tabcolsep}{3pt}
\renewcommand{\arraystretch}{0.9}
\resizebox{\columnwidth}{!}{%
\begin{tabular}{@{}llrr@{\hspace{5pt}}|@{\hspace{5pt}}llrr@{}}
\toprule
Method & Best config & Acc. & Gap 
& Method & Best config & Acc. & Gap \\
\midrule
DARE+avg & $p{=}0.9$ & 59.2 & 9.1 
& DARE+TIES$^\dagger$ & $p{=}0.9,k{=}0.2$ & 57.9 & 10.4 \\

Task arith. & $\lambda{=}1.0$ & 59.0 & 9.3 
& DELLA$^\dagger$ & $p{=}0.9,k{=}0.2$ & 55.3 & 13.0 \\

Simple avg & --- & 58.8 & 9.5 
& TIES$^\dagger$ & $k{=}0.5$ & 53.0 & 15.3 \\
\bottomrule
\end{tabular}%
}
\vspace*{-0.5cm}
\end{table}

\noindent\textbf{Asymmetric composition gap.}
The overall composition gap is 9.4\% (95\% CI: 9.0, 9.8). The per-group breakdown (Fig.~\ref{fig:asymmetric_gap}) shows an asymmetry. The three bird groups (G1, G2, G3), which dominate the training data, lose accuracy. The two minority groups (G4, G5) gain accuracy. 
\\
This follows from Eq.~\ref{eq:displacement}. Under $1/N$ averaging, each group receives equal weight regardless of task vector magnitude. Joint training, dominated by passerine data, allocates disproportionate capacity to large groups. Merging reverses this bias. Norm-adjusted weighting does not improve performance. Uniform weighting outperforms at every scaling coefficient (ablation on Appendix~\ref{app:norm_ablation}). This is consistent with near-orthogonality.


\noindent\textbf{Probing vs.\ local structure.}
$k$-NN ($k{=}1$) yields a composition gap of only 2.3\% (joint: 75.1\%, merged: 72.8\%), $4\times$ reduction compared to linear probing (9.4\%), indicating that local feature structure is largely preserved after merging but global subspace arrangement shifts enough to penalize linear classification.
\vspace*{-0.2cm}
\begin{tcolorbox}[colback=gray!10, colframe=black, boxrule=1.0pt, arc=1mm,left=0.5mm, right=0.5mm, top=0.5mm, bottom=0.5mm]
\centering
{\small{\textbf{F2:} 
Composition redistributes capacity from majority to minority taxa. Local feature structure is preserved; gap is driven by global subspace arrangement.
}}
\end{tcolorbox}
\vspace*{-0.3cm}
\subsection{Experiment 3: Regional Composition}
\label{sec:regional}
\vspace*{-0.2cm}
We test whether near-orthogonality holds for geographic as well as taxonomic grouping, and whether composition enables zero-shot transfer to new regions. Four regional models are fine-tuned (R1--R4).
We compose them using uniform, ecological, and incremental strategies. Cross-region leave-one-out (merge three regions, evaluate on fourth) tests zero-shot transfer.
\\
Regional task vectors show higher cosine values, ranging from 0.083 to 0.116 (Table~\ref{tab:regional_geom}). This is expected, as all models classify birds. The extends to a second biological axis: cross-taxa (0.01–0.04) $<$ intra-avian (0.08–0.09) $<$ regional (0.08–0.12).


\begin{table}[t]
\centering
\caption{Regional task vector geometry. All pairs show near-chance sign agreement.}
\label{tab:regional_geom}
\vspace*{-0.2cm}
\small
\setlength{\tabcolsep}{3pt}
\renewcommand{\arraystretch}{0.9}
\begin{tabular}{@{}lcc@{\hspace{10pt}}|@{\hspace{10pt}}lcc@{}}
\toprule
Pair & Cos. & Sign & Pair & Cos. & Sign \\
\midrule
R1--R2 & 0.116 & 0.548 & R1--R4 & 0.083 & 0.534 \\
R1--R3 & 0.104 & 0.543 & R2--R4 & 0.084 & 0.534 \\
R2--R3 & 0.113 & 0.545 & R3--R4 & 0.084 & 0.534 \\
\bottomrule
\end{tabular}
\vspace*{-0.2cm}
\end{table}

\noindent Uniform merging achieves 60.8\% accuracy across all four regions, a gap of 6.5\% from the jointly-trained baseline of 67.2\%. TIES again performs worst at 55.8\%, consistent with our previous findings. In leave-one-out evaluation, merging three regional task vectors produces an encoder that classifies species in the held-out region at 90.8\% of the accuracy of a dedicated single-region model. This occurs despite minimal species overlap between regions (Jaccard $< 0.034$). The merged model generalizes to species it has never seen during training, indicating that composition captures shared acoustic structure rather than memorizing species-specific features.
\vspace*{-0.3cm}
\subsection{Experiment 4: Domain Negation}
\label{sec:negation}
\vspace*{-0.2cm}

\begin{figure}[t]
\centering
\includegraphics[width=0.9\columnwidth]{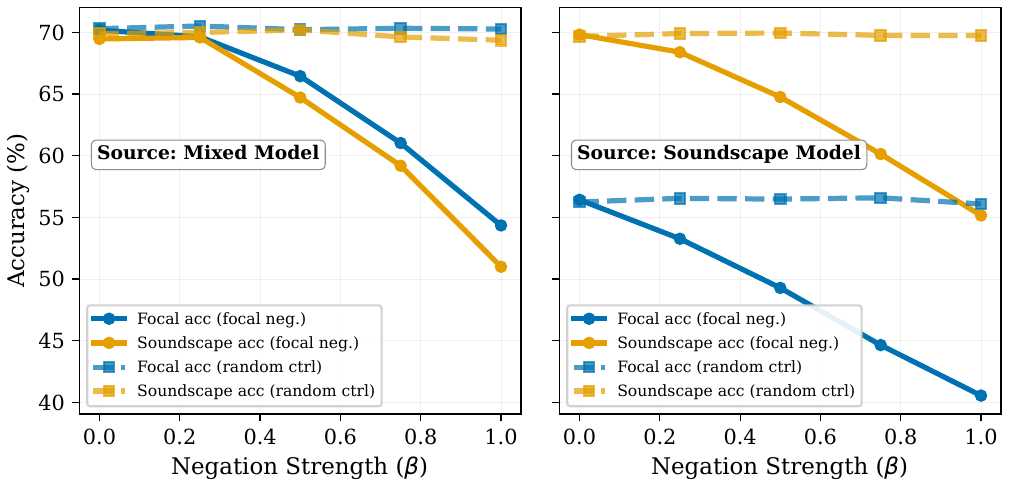}
\vspace*{-0.3cm}
\caption{Domain negation: accuracy vs.\ subtraction strength $\beta$ for focal negation (solid) and random-vector control (dashed). 
}
\label{fig:domain_negation}
\vspace*{-0.5cm}
\end{figure}

Task arithmetic allows subtraction as well. Subtracting a focal-recording task vector from a source model: $\boldsymbol{\theta}_{\mathrm{new}} = \boldsymbol{\theta}_{\mathrm{source}} - \beta \boldsymbol{\tau}_{\mathrm{focal}}$ should, in principle, remove focal characteristics and improve performance on soundscape data. However, domain negation fails because the `focal' recording is not a separable style but a canonical representation of species identity. Subtracting $\boldsymbol{\tau}_{\mathrm{focal}}$ thus erases core species information rather than removing a stylistic attribute. 
We test this by subtracting $\boldsymbol{\tau}_{\mathrm{focal}}$ from both mixed-domain and soundscape-only source models across $\beta$ (0--1.0), observing monotonic degradation of both focal and soundscape accuracy (Fig.~\ref{fig:domain_negation}). A random-vector control produces essentially no effect ($\pm 0.6\%$), confirming that degradation is direction-specific and stems from the entanglement of domain and identity in the encoder's weight space.
\begin{tcolorbox}[colback=gray!10, colframe=black, boxrule=1.0pt, arc=1mm,left=0.5mm, right=0.5mm, top=0.5mm, bottom=0.5mm]
\centering
{\small{\textbf{F3:} Composition works for taxonomic and geographic tasks with zero-shot transfer, but fails for domain negation because recordings may entangle with species identity.}}
\end{tcolorbox}

\noindent Additional experiments 
are in the Supplementary Material.

\vspace*{-0.2cm}
\section{Discussion}
\label{sec:discussion}
\vspace*{-0.1cm}

\textbf{Toward collaborative model building in bioacoustics.}
Current bioacoustic monitoring relies on either monolithic classifiers (BirdNET, Perch) that require centralized retraining for updates, or isolated specialists that cannot share knowledge. Task arithmetic offers an alternative. Each research group trains on its own data, contributes a task vector, and the community assembles multi-taxa classifiers through weight-space arithmetic. Coordination cost is minimal. By agreeing on a base model, adding a new taxon requires no access to existing training data. The observed asymmetric property (F2) suggests this method may be particularly relevant for conservation programs mandated to monitor across taxa, not just within dominant groups.
\\
\textbf{One-shot federated learning with a favorable inductive bias.}
Tao et al.~\cite{tao2024oneshot} show that task arithmetic is mathematically equivalent to one-shot FedAvg. In federated learning, data heterogeneity causes drift degrading merged models~\cite{karimireddy2020scaffold}. Bioacoustic composition inverts this: spectral heterogeneity produces orthogonal task vectors. The gradient from cross-taxa to intra-avian vectors aligns with the acoustic niche hypothesis.

\vspace*{-0.2cm}
\section{Conclusion}
\vspace*{-0.1cm}

Our results affirmatively answer both questions posed in~\S\ref{sec:intro}, as summarized in findings F1--F3. Acoustic niche partitioning extends to weight space: taxa in distinct spectral bands produce modular, composable parameter updates. This geometric regime distinguishes bioacoustics from vision and explains why conflict-resolution methods are counterproductive here. The association (F1) further suggests that composition quality can be assessed \textit{a priori} from acoustic properties alone. Future work includes validating this criterion across architectures, scaling composition to finer taxonomic resolution where orthogonality may weaken, and designing classifiers that exploit preserved local feature structure (F2) to narrow composition gap. As monitoring expands to diverse taxa and under-surveyed regions, task vector arithmetic offers the community a path toward shared multi-taxa classifiers: contributing learned knowledge as ecological coverage grows, without centralizing data or retraining.

\clearpage

\bibliographystyle{IEEEtran}
\bibliography{mybib}
\vspace{2cm}

\noindent\textbf{Generative AI Use Disclosure}\\
Generative AI tools were used for editing prose and debugging code analysis. All scientific content, experimental design, implementation, and interpretation of results are the authors' own work.

\clearpage

\appendix
\setcounter{table}{0}
\renewcommand{\thetable}{S\arabic{table}}
\setcounter{figure}{0}
\renewcommand{\thefigure}{S\arabic{figure}}
\setcounter{equation}{0}
\renewcommand{\theequation}{S\arabic{equation}}

\twocolumn[
\begin{center}
\large{\textbf{Supplementary Material}\\\textit{Ecologically-Constrained Task Arithmetic for Multi-Taxa Bioacoustic\\
Classifiers Without Shared Data}}
\vspace{1.0cm}
\end{center}
]
\section{Motivating Deployment Scenarios}
\label{app:scenarios}

The problem setting in \S\ref{sec:setting} abstracts away why training data is unavailable. Here we describe scenarios that reduce to this formulation, clarifying the practical scope of task vector composition for bioacoustic monitoring.

\begin{enumerate}
    \item \textbf{Models without data.} A research group publishes a fine-tuned BEATs encoder for cetacean calls, but the recordings are not released due to data agreements or sensitive location information about endangered species. The checkpoint is available; the training data is not. This is already common: BirdNET~\cite{kahl2021birdnet} and Perch~\cite{williams2024perch} distribute only model weights. A practitioner wanting to combine a published cetacean model with their own passerine classifier faces similar to our problem: two specialists from the same base model, no access to original datasets.
    \item \textbf{Data too large to centralize.} Passive acoustic monitoring stations terabytes of audio per station per year (e.g., $\sim$3TB annually for continuous 48kHz WAV recording). A network of stations may accumulate petabytes within a few seasons. Downloading and jointly training on this volume is infeasible for most groups. However, each station \textit{can} fine-tune locally and export a task vector, always $d$ parameters (${\sim}$360\,MB for BEATs), regardless of dataset size. Composition converts a terabyte-scale logistics problem into a megabyte-scale model exchange.
    \item \textbf{Temporal mismatch.} For example, a bird classifier covering 577 species is deployed in January. In June, a new amphibian dataset covering 63 species becomes available. Adding amphibian coverage via joint training requires re-accessing all original bird data and retraining from scratch. This coupling does not match the incremental nature of ecological data collection. Task vector composition decouples the update: the amphibian group trains a specialist, exports its task vector $\boldsymbol{\tau}_{\mathrm{amphibian}}$, and the merged model is reconstructed in under a second.
    \item \textbf{Geographic expansion without local data.} A monitoring program trained on East African and South Asian bird data needs to deploy in the Neotropics, where no local training data exists. If existing regional models capture transferable acoustic structure, composing their task vectors may bootstrap a classifier for the new region. Our leave-one-out experiments test this scenario (\S\ref{sec:regional}), finding that merged regional models classify species in the held-out region at 90.8\% of dedicated performance despite minimal species overlap.
\end{enumerate}


\section{Background: Model Merging Methods}
\label{app:merging_background}
\begin{figure}[t]
\centering
\includegraphics[width=0.95\columnwidth]{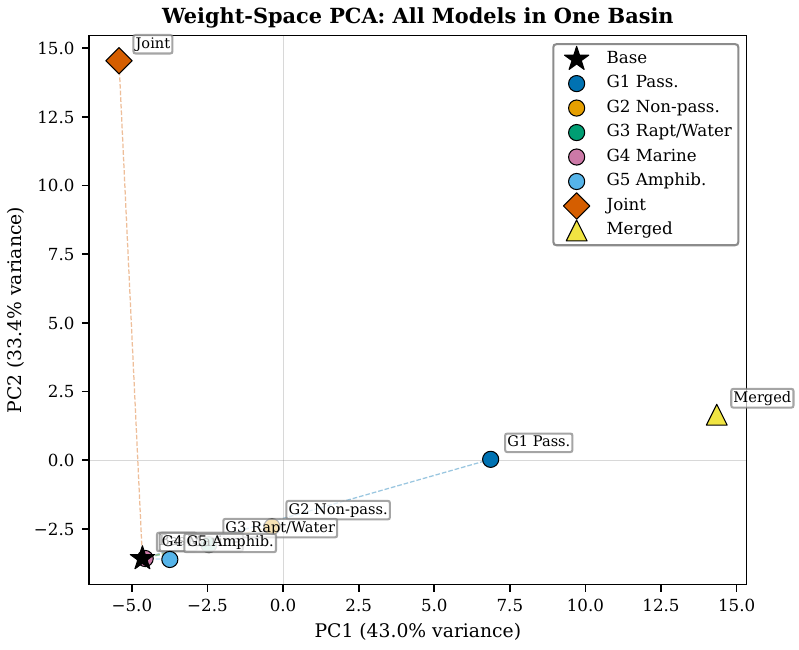}
\vspace*{-0.3cm}
\caption{PCA of models checkpoints in encoder weight space. 
}
\label{fig:weight_pca}
\vspace*{-0.5cm}
\end{figure}
We provide extended background on the merging methods evaluated. 

\subsection{Task arithmetic and linear mode connectivity}

Task arithmetic \cite{ilharco2023editing} relies on the observation that
models fine-tuned from a shared initialization tend to remain in the same loss basin. This property, known as linear mode connectivity (LMC) \cite{frankle2020lmc}, means that for any two models $\boldsymbol{\theta}_A$ and $\boldsymbol{\theta}_B$ fine-tuned from $\boldsymbol{\theta}_0$, the linear interpolation $\boldsymbol{\theta}(\alpha) = \alpha \boldsymbol{\theta}_A + (1-\alpha)\boldsymbol{\theta}_B$ does not encounter a loss barrier; the loss along the path never exceeds the loss at either endpoint. When LMC holds, averaging weights is geometrically safe because the merged model remains in the same basin as the specialists.
\\
The task vector $\boldsymbol{\tau}_i = \boldsymbol{\theta}_i -
\boldsymbol{\theta}_0$ captures the direction and magnitude of changes from fine-tuning. Adding task vectors ($\boldsymbol{\theta}_0 + \sum_i
\boldsymbol{\tau}_i$) composes knowledge, while subtracting a vector
($\boldsymbol{\theta} - \boldsymbol{\tau}_i$) removes it. In vision, adding task vectors across eight classification tasks produced monotonic multi-task improvement \cite{ilharco2023editing}. However, most tests have used CLIP ViT models; audio applications remain limited and recent \cite{rittergutierrez2025audio}.
\\
Figure~\ref{fig:weight_pca} confirms LMC: a PCA projection of encoder weights places all specialists, the jointly-trained model, and the merged model in a single basin around the pretrained checkpoint. The first two principal components capture 76.4\% of variance. Distance from the base tracks task vector L2 norm: G1 (passerines, $||\boldsymbol{\tau}|| = 13.58)$ is farthest, G4 (marine mammals, $||\boldsymbol{\tau}|| = 1.45$) nearest. Merged and joint models are close but not coincident, geometrically visualizing the composition gap.
\subsection{TIES-Merging}
TIES~\cite{yadav2023ties} addresses \textit{sign conflicts}, parameters where different task vectors disagree on the direction of change from
$\boldsymbol{\theta}_0$. TIES operates in three steps:

\begin{enumerate}
    \item Trim: Zero out the bottom $k$ fraction of parameters by magnitude in each task vector, retaining only the largest changes.
    \item Elect sign: For each parameter, take a majority vote
    across task vectors to determine the dominant sign.
    \item Disjoint merge: Average only the values that agree with the elected sign; discard conflicting values.
\end{enumerate}

\subsection{DARE}
DARE \cite{yu2024dare} builds on the observation that fine-tuning deltas are highly redundant: dropping most individual parameter changes often leaves the merged model's behavior unchanged. DARE randomly zeros each parameter with probability $p$ and rescales the survivors by $1/(1-p)$ to preserve the expected value. This is similar to applying dropout in weight space rather than activation space. At high drop rates ($p \geq 0.9$), task vectors become extremely sparse, which can reduce interference during merging. The $1/(1-p)$ rescaling factor requires FP32 precision to avoid numerical instability at high drop rates.

\subsection{DELLA}
DELLA \cite{panigrahi2024della} modifies DARE by using magnitude-proportional dropout instead of uniform dropout: parameters with larger absolute values have a lower drop probability, which preferentially preserves the largest weight changes. The drop probability for parameter $i$ is $p_i = (1 - |\delta_i|/\max(|\delta|)) \cdot p_{\mathrm{target}}$. After sparsification, TIES sign election is applied. In our experiments, DELLA underperforms simple averaging because preserving larger magnitudes does not overcome the destructive effects of the subsequent sign election step.


\section{Extended Dataset Details}
\label{app:datasets}

\subsection{Source datasets}

Table~\ref{tab:source_datasets} provides full details on the six source datasets before preprocessing and group assignment.

\subsection{Species group construction}

Groups G1--G3 are defined by partitioning all bird species from BirdCLEF 2023, 2024, and 2025 according to eBird and IOC taxonomic orders:

\begin{itemize}
    \item \textbf{G1 Passerines}: Order Passeriformes (336 species). The largest and most acoustically diverse group, spanning songbirds from three continents.
    \item \textbf{G2 Non-passerine birds}: Orders Strigiformes, Piciformes, Columbiformes, Cuculiformes, Caprimulgiformes, Apodiformes, Trogoniformes, Coraciiformes, Bucerotiformes, Galbuliformes, Psittaciformes (157 species).
    \item \textbf{G3 Raptors and waterbirds}: Orders Accipitriformes, Falconiformes, Charadriiformes, Anseriformes, Pelecaniformes, Suliformes, Ciconiiformes, Gruiformes, Podicipediformes, Phoenicopteriformes, Procellariiformes (84 species).
    \item \textbf{G4 Marine mammals}: Cetaceans and pinnipeds from the Watkins database (21 species). Recorded via hydrophones in ocean environments---a categorically different recording medium from terrestrial microphones.
    \item \textbf{G5 Amphibians}: Neotropical anurans from AnuraSet plus amphibian species from BirdCLEF 2025 (63 species).
\end{itemize}

The grouping criterion is biological taxonomy rather than acoustic clustering. This choice tests whether ecological relationships, which correlate with acoustic properties but are not identical to them, predict task vector geometry.

\subsection{Split statistics}

Table~\ref{tab:split_stats} reports the train, validation, test, and zero-shot split sizes for each group.

\begin{table}[!htbp]
\centering
\caption{Per-group split statistics. Zero-shot: clips from species held out entirely (not used in this paper but available for future work).}
\label{tab:split_stats}
\small
\resizebox{\linewidth}{!}{
\begin{tabular}{lrrrrr}
\toprule
Group & Classes & Train & Val & Test & Zero-shot \\
\midrule
G1 Passerines & 336 & 80{,}761 & 11{,}443 & 22{,}966 & 2{,}028 \\
G2 Non-pass. & 157 & 37{,}574 & 5{,}312 & 10{,}674 & 3{,}498 \\
G3 Rapt./Wat. & 84 & 20{,}709 & 2{,}935 & 5{,}890 & 3{,}915 \\
G4 Marine m. & 21 & 1{,}402 & 188 & 385 & 1{,}396 \\
G5 Amphibians & 63 & 11{,}898 & 1{,}682 & 3{,}379 & 3{,}417 \\
\midrule
Union & 661 & 152{,}344 & 21{,}560 & 43{,}294 & --- \\
\bottomrule
\end{tabular}
}
\end{table}

\subsection{Regional datasets for Experiment 3}

Each BirdCLEF competition corresponds to a geographic region. BirdCLEF 2025
is multi-taxa; only Aves species are included in R3 (non-bird taxa are
assigned to species groups G4/G5 instead). BirdSet POW~\cite{rauch2024birdset} provides a fourth
region (North America) from soundscape recordings at Powdermill Nature
Reserve, Pennsylvania.

Species overlap between regions is minimal: pairwise Jaccard similarity
ranges from 0.008 to 0.034, confirming that the regional datasets cover
largely non-overlapping species assemblages. This near-disjointness makes the
leave-one-out transfer result notable: the merged model classifies species at
90.8\% of dedicated performance despite never encountering those species
during any fine-tuning run.

\subsection{Domain negation datasets for Experiment 4}

Experiment 4 requires three domain-specific datasets drawn from the same
species pool: (i)~a \textit{focal} set of clean, close-range Xeno-canto
recordings (the BirdCLEF training data itself); (ii)~a \textit{soundscape}
set of continuous field recordings with background noise, overlapping species,
and variable signal-to-noise ratios; and (iii)~a \textit{mixed} set combining
both. The focal set is defined by the BirdCLEF source recordings. The
soundscape set is drawn from BirdSet evaluation soundscapes, filtered to
species that overlap with the focal set.

\subsection{Note on recording medium}

Group G4 (marine mammals) was recorded with hydrophones in underwater environments, while all other groups used terrestrial microphones. This introduces a categorical difference in recording medium that cannot be separated from the taxonomic difference. As noted in the main text, the near-zero cosine similarity between G4 and all other groups may reflect recording environment as much as acoustic niche separation. However, intra-avian cosine values for G1--G3 pairs, all recorded with terrestrial microphones, are substantially below vision benchmarks. The graded ordering, with intra-avian similarity higher than bird-amphibian similarity, which is in turn higher than bird-marine mammal similarity, is consistent with an acoustic distance interpretation.

\begin{table}[!htbp]
\centering
\caption{Training hyperparameters (fixed across all runs).}
\label{tab:hyperparams}
\small
\begin{tabular}{ll}
\toprule
Parameter & Value \\
\midrule
\multicolumn{2}{l}{\textit{Architecture}} \\
Base model & BEATs iter3+ AS2M (90M params) \\
Encoder output & 768-dim, mean-pooled over time \\
Classifier head & Linear(768, $n_{\mathrm{classes}}$), Xavier init \\
\midrule
\multicolumn{2}{l}{\textit{Optimization}} \\
Optimizer & AdamW \\
Learning rate & $1 \times 10^{-5}$ \\
Weight decay & 0.01 \\
LR schedule & OneCycleLR (cosine + linear warmup) \\
Warmup steps & 500 \\
Batch size & 32 \\
Max epochs & 20 \\
Early stopping & Patience 5 (val loss) \\
Gradient clipping & Max norm 1.0 \\
Precision & BF16 mixed precision \\
\midrule
\multicolumn{2}{l}{\textit{Regularization}} \\
Label smoothing & $\varepsilon = 0.1$ \\
Mixup & $\alpha = 0.3$ (applied to 50\% of batches) \\
SpecAugment & 2 time masks ($T{=}50$ samples each) \\
Encoder freeze & First 2 epochs (head-only warmup) \\
\midrule
\multicolumn{2}{l}{\textit{Audio}} \\
Sample rate & 16{,}000\,Hz \\
Clip duration & 5.0\,s (80{,}000 samples) \\
Spectrogram & 128 mel bins, 25\,ms window, 10\,ms hop \\
\bottomrule
\end{tabular}
\end{table}

\begin{table*}[tb]
\centering
\caption{Source datasets before preprocessing. Clip counts reflect post-preprocessing (16\,kHz mono, 5\,s, $-$60\,dB energy filter).}
\label{tab:source_datasets}
\small
\begin{tabular}{llllrrl}
\toprule
Dataset & Region & Taxa & Recording type & Species & Raw clips & Access \\
\midrule
BirdCLEF 2023 & East Africa & Aves & Focal (Xeno-canto) & 264 & 142{,}923 & Kaggle \\
BirdCLEF 2024 & South Asia & Aves & Focal (Xeno-canto) & 182 & 209{,}236 & Kaggle \\
BirdCLEF 2025 & Neotropics & Multi-taxa$^a$ & Focal (Xeno-canto) & 206 & 210{,}855 & Kaggle \\
BirdSet POW & N.\ America & Aves & Soundscape & 48 & 166{,}536 & HuggingFace \\
Watkins & Global oceans & Mammalia & Focal (hydrophone) & 31$^b$ & 3{,}881 & WHOI \\
AnuraSet & Neotropics & Amphibia & Soundscape & 42 & 141{,}039 & Zenodo \\
\bottomrule
\end{tabular}\\[3pt]
\raggedright\footnotesize $^a$BirdCLEF 2025 includes Aves, Insecta, Amphibia, Mammalia. Only Aves used for regional experiments (R3); Amphibia in AnuraSet for G5; Insecta dropped (3 species, 111 clips-insufficient). 9 terrestrial Mammalia species (raccoons, jaguars, sloths) dropped:not marine, not fitting any group.\\
$^b$Watkins contains 60+ species; after preprocessing, 31 species met minimum sample thresholds. Final G4 uses 21 species after the 500-clip cap.
\end{table*}

\section{Extended Training Protocol}
\label{app:training}

\subsection{Full hyperparameter specification}

Table~\ref{tab:hyperparams} reports the complete hyperparameter configuration. The same settings were implied across all fine-tuning runs via config hashing with SHA-256 prefix \texttt{c4c3cf3b}.

\subsection{Encoder freeze schedule}

For the first 2 epochs, only the classification head parameters are trainable (16{,}149 parameters for G4 with 21 classes, up to 259{,}104 for G1 with 336 classes). The encoder's 90M parameters remain frozen. Freezing the encoder allows the randomly initialized head to reach a statble state before encoder gradients begin flowing. This prevents large gradients early in training from pushing the encoder far from $\boldsymbol{\theta}_0$, which is important for maintaining linear mode connectivity.

At epoch 3, the full encoder is unfrozen and the optimizer and scheduler are recreated to include encoder parameters. The learning rate ($10^{-5}$) and weight decay ($0.01$) are set conservatively to limit deviation from the pretrained basin.




\subsection{Task vector computation}

Task vectors are computed by subtracting $\boldsymbol{\theta}_0$ from each specialist's encoder weights:
\[
    \boldsymbol{\tau}_i = \{(\boldsymbol{\theta}_i^{(\ell)} - \boldsymbol{\theta}_0^{(\ell)}) \mid \ell \in \text{encoder layers}\}
\]
This covers all 250 encoder parameter tensors, including attention weights, layer norms, and feed-forward layers. The classification head is excluded because its dimensions are incompatible across groups. The resulting task vector has the same number of parameters as the encoder, approximately 90 million, and requires about 360 MB of storage in FP32.

\section{Supplementary Experiments}
\label{app:supplementary_experiments}

\subsection{Data efficiency}
\label{app:data_efficiency}

We test whether task vector composition remains robust when contributors have limited data. To do this, we fine-tuned G4 (marine mammals) on subsets of its training data (25\%, 50\%, 75\%, and 100\%) and merged the resulting task vectors into the full five-group system.

Table~\ref{tab:data_eff} shows the results. Even with only 25\% of the training data (341 clips), the merged system achieves 93.3\% accuracy on G4 and 58.8\% accuracy across all groups. These numbers are essentially identical to those obtained with the full dataset (93.8\% and 58.8\%). The task vector direction converges rapidly: cosine similarity between the subsampled and full-data task vectors jumps from 0.035 at 25\% to 0.78 at 50\% and 0.91 at 75\%. The direction of fine-tuning is established early; additional data primarily increases magnitude.

\begin{table}[!htbp]
\centering
\caption{Data efficiency: G4 marine mammals fine-tuned on data subsets, then
merged into full 5-group system.}
\label{tab:data_eff}
\small
\begin{tabular}{rrrrc}
\toprule
Data \% & Clips & G4 acc & All acc & $\cos(\boldsymbol{\tau}_{x\%},
\boldsymbol{\tau}_{100\%})$ \\
\midrule
25\% & 341 & 93.3\% & 58.8\% & 0.035 \\
50\% & 696 & 93.3\% & 58.9\% & 0.78 \\
75\% & 1{,}043 & 93.3\% & 59.1\% & 0.91 \\
100\% & 1{,}402 & 93.8\% & 58.8\% & 1.00 \\
\bottomrule
\end{tabular}
\end{table}

In practice, this means that low-resource taxa, such as poorly studied marine mammals or rare amphibians, can contribute effectively to a merged system even with limited training data. The task vector direction, which determines where the merged model sits in weight space, stabilizes well before magnitude convergence.

\subsection{Per-layer task vector structure}
\label{app:per_layer}

\begin{figure*}[t]
\centering
\includegraphics[width=0.75\textwidth]{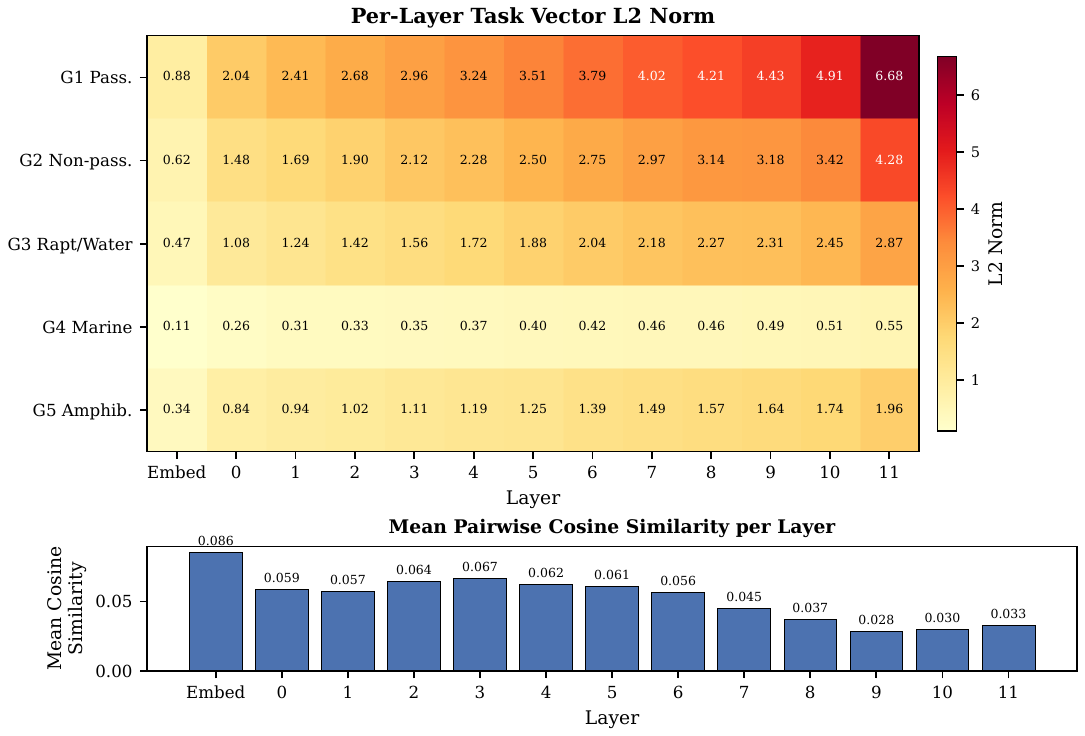}
\caption{Per-layer task vector structure. \textbf{Top}: L2 norm per group
per layer; all groups increase toward later layers, with G1 (passerines)
dominating. \textbf{Bottom}: mean pairwise cosine similarity per layer;
similarity decreases toward later layers, reaching 0.028 at layer~9.
Later layers modify more but in more orthogonal directions.}
\label{fig:layer_heatmap}
\end{figure*}

Figure~\ref{fig:layer_heatmap} shows a layer-wise decomposition of task vector geometry across the 12 transformer layers and the embedding layer. L2 norms increase monotonically from the embedding layer to layer 11 for all groups (e.g., G1: 0.88\,$\to$\,6.68). This indicates that later layers undergo the largest changes during fine-tuning. Mean pairwise cosine similarity follows the opposite trend, decreasing from 0.086 at the embedding layer to 0.028--0.030 at layers 9 and 10. The layers that change most do so in the most orthogonal directions. This divergence, where large modifications produce minimal interference, is the per-layer mechanism underlying the near-orthogonal global geometry reported in the main text. The pattern supports the hypothesis that early layers encode shared acoustic primitives such as spectral edges and onset detection, while later layers specialize for taxon-specific call structure.

\vspace*{-.4cm}
\subsection{Continual learning}
\label{app:continual}

We simulate adding a new taxonomic group to a deployed system by starting
with a G1--G4 encoder and incorporating G5 (amphibians). Table~\ref{tab:continual} compares five strategies.

\begin{enumerate}
    \item \textbf{Baseline}: The existing G1--G4 merged encoder evaluated
    on all five groups without modification.
    \item \textbf{Joint retraining}: Retrain on all G1--G5 data from scratch, which requires full data access and serves as an upper bound.
    \item \textbf{Fine-tune joint}: Fine-tune the existing G1--G4 joint
    model on G5 data, with a sweep
    over learning rates $\{10^{-5}, 5{\times}10^{-6}, 10^{-6}\}$.
    \item \textbf{Merge joint TV}: Compute
    $\boldsymbol{\tau}_{\mathrm{G5}}$ independently and add it to the
    G1--G4 joint model's task vector:
    $\boldsymbol{\theta}_{\text{base}} + \tfrac{1}{2}(\boldsymbol{\tau}_{\text{G1--G4}} + \boldsymbol{\tau}_{\text{G5}})$.
    \item \textbf{Merge individual TVs}: Compute
    $\boldsymbol{\tau}_{\mathrm{G5}}$ independently and average it task vectors:
    $\boldsymbol{\theta}_{\text{base}} + \tfrac{1}{5}\sum_{i=1}^{5}\boldsymbol{\tau}_{\text{G}i}$.
\end{enumerate}

\begin{table}[!htbp]
\centering
\caption{
Continual learning: adding G5 (amphibians) to an existing G1--G4 system. \emph{Old Mean} = mean accuracy across G1--G4 species; \emph{Forget} = drop relative to baseline; 
}
\label{tab:continual}
\small
\resizebox{\linewidth}{!}{
\begin{tabular}{lccc}
\toprule
\textbf{Strategy} & \textbf{Old Mean (\%)} & \textbf{G5 Acc (\%)} & \textbf{Forget (pp)} \\
\midrule
Baseline (G1--G4 only)       & 77.6 & 61.6 & ---    \\
Joint retraining (G1--G5)    & 78.0 & 61.6 & $-$0.4 \\
Fine-tune joint on G5        & 76.2 & 66.4 & +1.3   \\
Merge joint TV               & 74.6 & 64.2 & +2.9   \\
Merge individual TVs         & 72.3 & 62.0 & +5.3   \\
\bottomrule
\end{tabular}
}
\end{table}

\begin{figure}[t]
\centering
\includegraphics[width=\columnwidth]{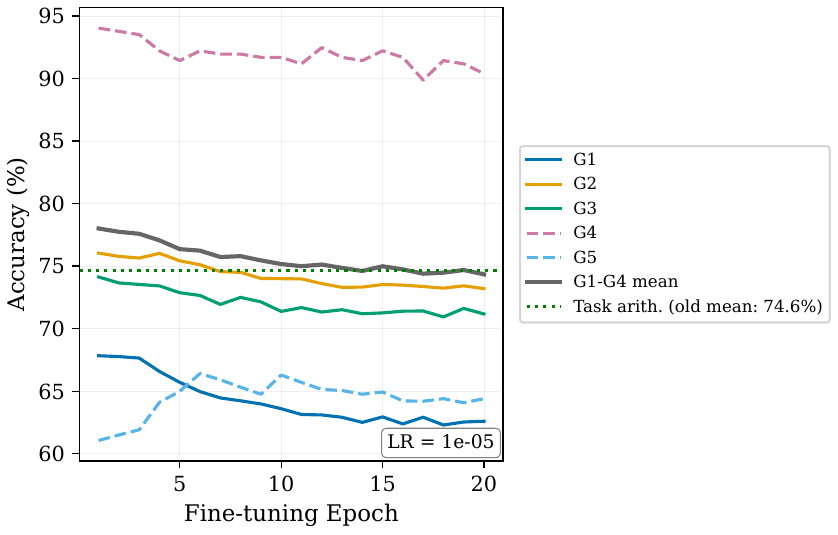}
\caption{Per-group accuracy during fine-tuning of the G1--G4 joint model
on G5 data (LR\,=\,$10^{-5}$). Dashed line: mean accuracy
achieved by task vector merging (no fine-tuning required).}
\label{fig:forgetting_dynamics}
\end{figure}

The two merging
strategies trade forgetting for practical benefits: merging the joint task
vector (+2.9\%) outperforms merging individual vectors (+5.3\%) because the joint encoder already captures inter-group structure. Notably, G4 (marine mammals) \emph{improves} after merging (92.5\%\,$\to$\,93.2\%),
consistent with the capacity redistribution observed in the main
experiments (Finding~2). In all cases, forgetting is bounded and
predictable via Equation~\ref{eq:displacement}. Merging's value proposition is therefore not knowledge preservation but computational efficiency and data privacy: a new contributor can extend the system's taxonomic coverage without accessing any existing training data.
\\
Figure~\ref{fig:forgetting_dynamics} shows per-group accuracy trajectories during fine-tuning with a learning rate of $10^{-5}$. G4 remains stable throughout. G1 declines steadily as the group most affected by forgetting, reflecting its large representation in the joint encoder. G5 accuracy rises sharply in the first five epochs before plateauing. The dashed line marks the merge-based alternative, which incurs a fixed accuracy penalty without requiring additional training.

\subsection{Group-level confusion analysis}
\label{app:confusion}

\begin{figure}[t]
\centering
\includegraphics[width=0.85\columnwidth]{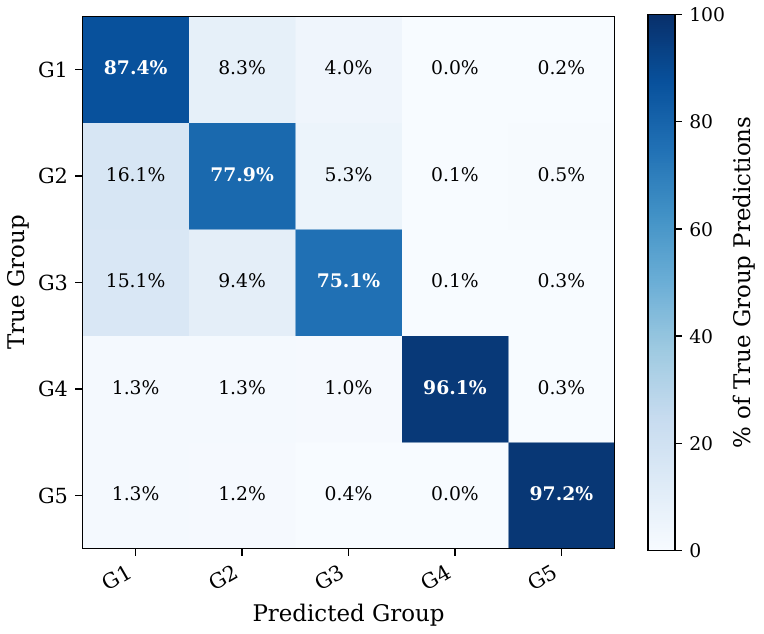}
\caption{Group-level confusion matrix for merged encoder 
}
\label{fig:group_confusion}
\vspace*{-.4cm}
\end{figure}

Figure~\ref{fig:group_confusion} shows that errors in the merged encoder concentrate almost entirely within the three bird groups. G2\,$\to$\,G1 (16.1\%), G3\,$\to$\,G1 (15.1\%), G1\,$\to$\,G2 (8.3\%). G4 (marine mammals, 96.1\%) and G5 (amphibians, 97.2\%) are nearly isolated, with $\leq$\,1.3\% misclassification to any other group. This pattern mirrors the cosine similarity structure: intra-avian pairs show the most task vector overlap and classification confusion, while cross-taxa pairs are both geometrically and functionally separable. The asymmetry between G1\,$\to$\,G2 and G2\,$\to$\,G1 reflects G1's larger task vector magnitude, which draws more representational weight in the merged encoder. These results indicate that the composition gap stems from intra-avian interference rather than cross-taxa misclassification, and suggest that further gains would require reducing overlap between bird subgroups.

\subsection{Compute efficiency}
\label{app:compute}

Table~\ref{tab:compute} compares the computational cost of task vector composition versus joint retraining.

\begin{table}[t]
\centering
\caption{Compute comparison: task vector merging vs.\ joint retraining.}
\label{tab:compute}
\small
\begin{tabular}{lrr}
\toprule
Operation & Merging & Joint \\
\midrule
Initial training & 4.87 GPU-h$^a$ & 4.67 GPU-h \\
Adding one new group & 0.38 GPU-h$^b$ + 0.76\,s & 4.67 GPU-h \\
\midrule
Update speedup & \multicolumn{2}{c}{${\sim}6\times$ (total wall-clock)} \\
Marginal merge cost & \multicolumn{2}{c}{0.76\,s (CPU arithmetic only)} \\
\bottomrule
\end{tabular}\\[3pt]
\raggedright\footnotesize $^a$Sum of five independent specialist training
runs (embarrassingly parallel).\\
$^b$Training one specialist on the new group only.
\end{table}

Initial training costs are comparable. Independent specialist training totals 4.87 GPU-hours (NVIDIA A100), but it is embarrassingly parallel because all five runs can execute simultaneously on separate GPUs. Joint training takes 4.67 GPU-hours in a single run. The difference appears when updating the system. Adding a new group via merging requires training one specialist (0.38 GPU-hours for G5-sized data) plus 0.76 seconds of CPU arithmetic, compared to 4.67 GPU-hours to retrain the joint model from scratch. This represents roughly a sixfold reduction in total wall-clock time. The ratio grows with the number of existing groups, since joint retraining cost scales with total dataset size while merging cost depends only on the new group. Given precomputed task vectors, reconstructing the full multi-taxa encoder takes under one second.

\subsection{Norm-adjusted weighting ablation}
\label{app:norm_ablation}

Task vector L2 norms vary by a factor of 9.4 across groups (G1: 13.58, G4: 1.45). This range raises the question of whether norm-adjusted weighting could reduce the composition gap. We test inverse-norm weighting
$w_i \propto ||\boldsymbol{\tau}_i||_2^{-\gamma}$ for
$\gamma \in \{0.3, 0.5, 0.7, 1.0, 1.5, 2.0\}$ (Table~\ref{tab:norm_ablation}).

\begin{table}[!htbp]
\centering
\caption{Norm-adjusted vs.\ uniform weighting across scaling factors
$\lambda$. $\Delta$ = uniform $-$ norm-adjusted. Joint baseline: 68.3\%.}
\label{tab:norm_ablation}
\small
\begin{tabular}{lcccccc}
\toprule
& \multicolumn{6}{c}{$\lambda$} \\
\cmidrule(lr){2-7}
\textbf{Weighting} & 0.3 & 0.5 & 0.7 & 1.0 & 1.5 & 2.0 \\
\midrule
Uniform       & 51.8 & 54.5 & 56.6 & 59.0 & \textbf{60.3} & 59.7 \\
Norm-adjusted & 50.1 & 51.6 & 52.5 & 54.5 & 56.5 & 57.1 \\
\midrule
$\Delta$      & +1.7 & +2.9 & +4.1 & +4.5 & +3.8 & +2.6 \\
\bottomrule
\end{tabular}
\end{table}

Uniform weighting outperforms norm-adjusted weighting at every scaling factor by 1.7 to 4.5\%. The best overall result is uniform weighting with $\lambda = 1.5$ at 60.3\%, while norm-adjusted peaks at $\lambda = 2.0$ with 57.1\%. This outcome is consistent with near-orthogonal geometry. When task vectors occupy largely non-overlapping parameter subsets, there is no interference to correct by equalizing magnitudes. The linear probe already adapts to differences in feature scale during its own training. 
\subsection{Merging method hyperparameter sensitivity}
\label{app:hyperparam}

For methods with tunable hyperparameters, we report full sweep results in Table~\ref{tab:hyperparam}.
\\
\noindent\textbf{Task arithmetic.} $\lambda = 1.0$ is optimal (59.0\%, equivalent to simple averaging with a scaling prefix). Performance degrades monotonically for $\lambda < 1.0$, consistent with the interpretation that the full task vector magnitude is informative rather than excessive.
\\
\textbf{DARE.} Performance remains stable across drop rates from $p = 0.5$ to $0.99$. Even at $p = 0.99$, where only 1\% of parameters survive, DARE matches simple averaging at 58.8\%. This stability reflects the high sparsity of bioacoustic task vectors: when most parameters are already near zero, random dropout has little left to remove.
\\
\textbf{TIES.} Performance falls below simple averaging at every tested trim fraction $k$. The best result is 53.0\% at $k = 0.5$, which is 5.8\% below simple averaging. This confirms that TIES failure is not a hyperparameter issue but a structural consequence of near-chance sign agreement, which ranges from 50.6\% to 53.9\%. At these levels, majority-sign resolution becomes effectively random.
\begin{table}[!htbp]
\centering
\caption{Merging method hyperparameter sweeps. Joint baseline: 68.3\%;
simple average: 58.8\%. Best result per method in bold.}
\label{tab:hyperparam}
\small
\begin{tabular}{llcc}
\toprule
\textbf{Method} & \textbf{Hyperparameter} & \textbf{Acc (\%)} & \textbf{Gap (pp)} \\
\midrule
\multirow{5}{*}{Task arithmetic}
  & $\lambda = 1.0$ & \textbf{59.0} & +9.5 \\
  & $\lambda = 0.8$ & 57.3 & +11.2 \\
  & $\lambda = 0.5$ & 54.3 & +14.2 \\
  & $\lambda = 0.3$ & 51.6 & +16.9 \\
  & $\lambda = 0.1$ & 48.8 & +19.6 \\
\midrule
\multirow{5}{*}{DARE + avg}
  & $p = 0.9$  & \textbf{59.2} & +9.3 \\
  & $p = 0.7$  & 59.0 & +9.5 \\
  & $p = 0.5$  & 58.9 & +9.6 \\
  & $p = 0.95$ & 58.9 & +9.6 \\
  & $p = 0.99$ & 58.8 & +9.7 \\
\midrule
\multirow{4}{*}{TIES}
  & $k = 0.5$ & \textbf{53.0} & +15.4 \\
  & $k = 0.8$ & 53.0 & +15.5 \\
  & $k = 0.2$ & 52.2 & +16.3 \\
  & $k = 0.1$ & 51.7 & +16.8 \\
\midrule
\multirow{2}{*}{DARE + TIES}
  & $p=0.9,\,k=0.2$ & \textbf{57.9} & +10.6 \\
  & $p=0.7,\,k=0.2$ & 55.3 & +13.2 \\
\midrule
\multirow{2}{*}{DELLA-TIES}
  & $p=0.9,\,k=0.2$ & \textbf{55.3} & +13.2 \\
  & $p=0.7,\,k=0.2$ & 53.7 & +14.8 \\
\bottomrule
\end{tabular}
\end{table}


\section{Spectral Distribution Distance Analysis}
\label{app:spectral}

We investigate the relationship between acoustic properties and task vector geometry (\S\ref{sec:composition}) by correlating pairwise spectral distribution distances with task vector cosine similarities across all 10 group pairs.

\subsection{Method}
For each of the five species groups, we randomly sample 500 training clips (or all clips if fewer than 500 are available). Each clip is converted to a 128-bin log-mel spectrogram (25\,ms window, 10\,ms hop, 16\,kHz). Averaging over time and across clips produces a single 128D mean log-mel energy profile per group. This profile serves as a ``spectral fingerprint'', where each group concentrates acoustic energy across frequency.
\\
We compute pairwise distances between the five profiles using three metrics. The primary metric is Jensen-Shannon divergence, which is symmetric, bounded, and robust to near-zero bins. For this metric, we convert log-mel profiles to probability distributions via softmax before computing divergence. The second metric is L2 distance on raw log-mel profiles, which captures both level and shape differences. The third metric is L2 distance after mean-centering, which removes overall recording level and isolates spectral shape. If the shape-only metric correlates as strongly as raw L2, then recording gain is not driving the result.
\\
Pairwise task vector cosine similarities come from the sparsity analysis (\S\ref{sec:composition}). We compute Spearman rank correlation between spectral distances and cosine similarities. Because the ten pairwise comparisons derive from only five group profiles and are not independent, we use a permutation test with 100,000 permutations rather than a parametric approximation. We recompute the correlation excluding the four pairs that involve G4, testing robustness to confound of hydrophone recording.

\subsection{Results}

Table~\ref{tab:spectral_corr} shows the correlation results. All three metrics yield strong negative correlations: groups with more different spectral content produce more orthogonal task vectors. The shape-only metric, with $\rho = -0.842$ and $p = 0.004$, confirms that spectral shape divergence, not recording gain, drives the relationship.

\begin{table}[t]
\centering
\caption{Spearman correlation between spectral distribution distance and task vector cosine similarity. Permutation $p$-values (100{,}000 permutations).}
\label{tab:spectral_corr}
\small
\begin{tabular}{lccc}
\toprule
Metric & $\rho$ & $p$ (perm.) & $n$ pairs \\
\midrule
JSD (primary) & $-$0.915 & 0.0006 & 10 \\
L2 (raw) & $-$0.867 & 0.002 & 10 \\
L2 (shape-only) & $-$0.842 & 0.004 & 10 \\
\midrule
JSD excl.\ G4 & $-$0.771 & 0.103 & 6 \\
\bottomrule
\end{tabular}
\end{table}

\noindent When we exclude G4-involved pairs, the negative trend persists with $\rho = -0.771$, but it does not reach significance at conventional thresholds. This reflects insufficient statistical power with only six data points rather than an absence of effect.
\\
Table~\ref{tab:spectral_pairs} shows the pair-level data, showing three distinct clusters. Bird--bird pairs (G1--G3; JSD 0.001--0.016) have near-identical spectral distributions and the highest cosine similarities. Bird--amphibian pairs (JSD 0.028--0.057) fall at intermediate distance and intermediate cosine (0.029--0.039). Pairs involving G4 have JSD values from 0.27 to 0.34, representing extreme spectral differences, and the lowest cosine similarities from 0.013 to 0.022. For these pairs, the hydrophone recording medium is an inherent confound that cannot be separated from the taxonomic difference. We therefore cannot definitively attribute the extreme spectral and geometric distances solely to biological acoustic niche partitioning.

\begin{table}[t]
\centering
\caption{Pairwise spectral distances and task vector cosine similarities. $\triangle$: pair involves hydrophone recordings (G4).}
\label{tab:spectral_pairs}
\small
\begin{tabular}{lccl}
\toprule
Pair & JSD & Cosine & Type \\
\midrule
G2--G3 & 0.001 & 0.093 & bird--bird \\
G1--G2 & 0.015 & 0.092 & bird--bird \\
G1--G3 & 0.016 & 0.085 & bird--bird \\
G1--G5 & 0.028 & 0.029 & bird--other \\
G2--G5 & 0.051 & 0.038 & bird--other \\
G3--G5 & 0.057 & 0.039 & bird--other \\
G4--G5$^\triangle$ & 0.266 & 0.022 & other--other \\
G2--G4$^\triangle$ & 0.300 & 0.013 & bird--other \\
G3--G4$^\triangle$ & 0.308 & 0.022 & bird--other \\
G1--G4$^\triangle$ & 0.337 & 0.014 & bird--other \\
\bottomrule
\end{tabular}
\vspace*{-0.6cm}
\end{table}

\noindent Two observations stand out. First, G2 (non-passerine birds) and G3 (raptors and waterbirds) have nearly identical spectral distributions (JSD $\approx$ 0.001), yet their task vectors remain near-orthogonal with cosine 0.093. Spectral distance is therefore a strong predictor of task vector geometry at the macro scale across clusters, but it does not fully explain within-cluster variation. Species-level identity contributes to weight-space geometry beyond the spectral envelope. Second, the scatter plot in the main text visualizes this three-cluster structure.

\section{Linear Probe Training Details}
\label{app:probe}

All linear probes are trained on frozen encoder features using Adam (lr $= 10^{-3}$), batch size 256, for 10 epochs. Each probe is a single linear layer that maps from the 768-dimensional BEATs embedding, averaged over the time axis, to the number of classes $n_{\mathrm{classes}}$. We apply no regularization, dropout, or data augmentation during probe training. This minimal classifier is designed to isolate encoder quality from probe complexity.
\\
For per-group probes, $n_{\mathrm{classes}}$ equals the number of species in that group (e.g., 336 for G1). For the unified probe, $n_{\mathrm{classes}} = 661$. Per-group probes assess how well the merged encoder preserves discriminability within each group, while the unified probe evaluates global multi-taxa classification performance. The composition gap is measured using the unified probe.
\\
We also evaluate using $k$-nearest neighbors with $k=1$. This approach requires no training: each test sample receives the label of its nearest neighbor in the training set, based on cosine distance in the encoder's 768-dimensional feature space. The $k$-NN diagnostic is non-parametric and insensitive to linear probe capacity.

\section{Reproducibility}
\label{app:reproducibility}

All fine-tuning runs are validated by a SHA-256 config hash (prefix \texttt{c4c3cf3b}) logged with each checkpoint. Runs with mismatched hashes are excluded.
Training uses seed 42 for data splitting and batch sampling. DARE and DELLA sparsification use per-vector seeds (base seed plus vector index) to ensure reproducibility while avoiding correlated masks.
Each checkpoint includes the config hash, per-epoch validation metrics, optimizer state, and the list of encoder parameter keys used for task vector computation (250 keys for BEATs). This enables post-hoc verification of identical architectures and training protocols across all models.

\end{document}